# Impact of Surface Passivation on the Efficiency and High-speed Modulation of III–V GaAs/AlGaAs Nanopillar Array LEDs

Bejoys Jacob,[1,2]* João Azevedo,[1,3] João Lourenço,[2] Filipe Camarneiro,[1] Jana B. Nieder,[1] Bruno Romeira[1]**

[1]INL – International Iberian Nanotechnology Laboratory, Av. Mestre José Veiga s/n, 4715-330, Braga, Portugal
[2]Centro-Ciências and LIP - Laboratório de Instrumentação e Física Experimental de Partículas, Departamento de Física, Faculdade de Ciências, Universidade de Lisboa, 1749-016 Lisboa, Portugal
[3]Departamento de Física, Universidade do Minho, Campus de Gualtar, 4710-057, Braga, Portugal
*bejoys.jacob@inl.int, ** bruno.romeira@inl.int

**Abstract**

III-V semiconductor nanolight sources with deep-subwavelength dimensions (<<1 µm) are essential for miniaturized photonic devices such as nanoLEDs and nanolasers. However, these nanoscale emitters typically suffer from substantial non-radiative recombination at room temperature, resulting in low efficiency and ultrashort lifetimes (<100 ps). Previous works have predominantly studied surface passivation of nanoLEDs under optical pumping conditions, while practical applications require electrically driven nanoLEDs. Here, we investigate the influence of surface passivation on the efficiency and high-speed modulation response of electrically pumped III-V GaAs/AlGaAs nanopillar array LEDs. Surface passivation was performed using ammonium sulphide chemical treatment followed by encapsulation with a 100 nm silicon nitride layer deposited via low-frequency plasma-enhanced chemical vapour deposition. Time-resolved electroluminescence (TREL) measurements reveal differential carrier lifetimes ($\tau$) of ~0.61 ns for nanoarray LEDs with pillar diameters of ~440 nm, a record-long lifetime for electrically driven GaAs-based nanopillar arrays. Under low injection conditions, the devices exhibited carrier lifetimes of ~0.41 ns, only 4-fold shorter than those of larger microLEDs ($\tau$~1.67 ns for 10 µm pillar diameter), indicating successful suppression of non-radiative effects and a low surface velocity, ranging from $S$~0.7 × 10$^4$ cm/s to 2.7 × 10$^4$ cm/s. This reveals a potential high internal quantum efficiency IQE~0.45 for our nanoLEDs operating under very high injection conditions, limited only by Auger recombination and self-heating effects at high current density. These miniaturized nanoLEDs with high radiative recombination efficiency and sub-nanosecond modulation response pave the way for optical data communications, energy efficient optical interconnects, AR/VR displays, and neuromorphic computing applications.

Keywords: III–V semiconductors, nanopillars, gallium arsenide, surface recombination, time-resolved electroluminescence, nanoLEDs

## Introduction

Miniaturized semiconductor LEDs are revolutionizing applications in lighting, displays, smartphones, automotive systems, augmented reality (AR), optical communication, and optical interconnects. Recently, inorganic microLEDs (2-50 µm) have become important in augmented and virtual reality (AR/VR) and in the Internet of Things (IoT), due to their high luminance, long lifetimes, and narrow pixel pitch.[1] Further scaling down nanoLEDs to sub-micron dimensions promises ultra-compact, energy-efficient, and high-bandwidth nano-optoelectronic devices. However, practical room-temperature nanoemitters face intrinsic limitations, such as substantial surface and Auger-related non-radiative recombination, particularly in III–V semiconductors, drastically reducing internal quantum efficiency (IQE).[2,3] Moreover, non-negligible dephasing processes can reduce the expected spontaneous emission rate enhancement (Purcell effect) in nanoLEDs.[4] Although optically pumped nanoscale emitters have demonstrated strong spontaneous emission enhancement with modulation speeds >50 GHz,[5,6] electrically driven nanoLEDs achieving both sub-ns modulation and high quantum efficiency have yet to be demonstrated.

Recent studies have shown electrically modulated, room-temperature operated nanoLEDs based on III-N, III-V, and 2D materials, aiming at sub-ns speeds and high efficiency. For example, III-V photonic crystal (PhC) nanoLEDs integrated with van der Waals heterostructures exhibited locally enhanced electroluminescence,[7] but the modulation speeds were limited to ~1 MHz. Other III-V PhC-based LEDs achieved sub-100 ps modulation speed, but showed extremely low external quantum efficiency (EQE~$10^{-5}$) at room-temperature.[8] Telecom-band single nanowire-LEDs on Si reached lifetimes of ~370 ps, yet provided only pW optical output.[9] An InP waveguide-coupled nanopillar (~350 nm) LED on Si using a metal-cavity design showed sub-200 ps lifetimes, nW power, and EQE~$10^{-4}$ at room temperature.[10] Such low optical powers (pW-nW range) and short carrier lifetimes reveal inherent limitations in single nanoLED devices. As a result, efforts have focused recently toward electrically pumped nanoarrays to improve optical output, including electrically driven GaN/InGaN QW nanowire array green lasers,[11] and core-shell GaN/InGaN nanowire-based LEDs with ~330 ps differential recombination lifetimes,[12] limited only by non-radiative recombination.

Recently, several passivation methods have been reported for III-V GaAs-based materials, highly relevant for near-infrared applications, with the goal of suppressing non-radiative recombination, including plasma passivation,[13] solution passivation with S- and N-containing chemicals,[14] use of $SiO_2$ sol-gel shell growth,[15] and epitaxial growth of a AlGaAs passivation layer.[16,17] Despite recent efforts, aside from a few theoretical works,[2,18,19] experimental studies simultaneously addressing efficiency and speed modulation in III-V GaAs-based nanoarray LEDs remain largely unexplored. In our recent work, we demonstrated a surface passivation method that led to a 29-fold increase of photoluminescence involving GaAs/AlGaAs dry-etched nanopillars treated with ammonium sulphide followed by encapsulation with $Si_xN_y$ deposited via low-frequency plasma enhanced chemical vapour deposition (PECVD).[3]

In this work, we provide a comprehensive investigation on the impact of this surface passivation treatment on the efficiency and modulation response of electrically-pumped III-V GaAs/AlGaAs *p-i-n* nanopillar array nanoLEDs encapsulated with $Si_xN_y$, forming both a passivation and an electrical insulating layer. Sulfurization prepares the initial surface for subsequent coating, while PECVD deposition effectively removes the native oxide from nanopillar sidewalls due to the enhanced ionic bombardment by $H^+$ ions at lower plasma frequencies (380 kHz). Time-resolved electroluminescence (TREL) measurements on nanopillars with diameters ranging from 440 nm to 870 nm reveal differential carrier lifetimes ranging from 0.41 ns to 0.61 ns, corresponding to record-long lifetimes for room temperature electrically driven III-V GaAs-based nanoLEDs operating in the near-infrared. We estimate a surface recombination velocity values ranging from $S$~$0.7 \times 10^4$ cm/s to $2.7 \times 10^4$ cm/s, indicating suppression of non-radiative effects in nanoarray LEDs. The experimental results are in a good agreement with simulations using a rate-equation model, from which we estimate IQE values ranging from 0.01 to 0.45 at low and high injection conditions, respectively, limited only by Auger recombination effect and self-heating effects at high current density.

**Design and fabrication**

The experimental study to evaluate the surface passivation treatment on the efficiency and high-speed modulation response employed electrically-pumped III-V AlGaAs/GaAs/AlGaAs nanopillar LEDs. Figure 1a shows a schematic of the 10×10 nanopillar array LED. The *p-i-n*-type III-V structure, Figure 1b, was grown by metalorganic chemical vapour deposition (MOCVD) on a semi-insulating (SI) GaAs substrate

(full epilayer details are provided in Supporting Information). The emitting intrinsic region consisted of a 280 nm-thick undoped bulk GaAs layer. The design included an AlAs/GaAs/AlAs double barrier quantum well (DBQW) region (~10 nm thick) on the *n*-type contact region, selected for investigating negative differential resistance effects in DBQW-based emitting devices,[20,21] which is beyond the scope of this work.

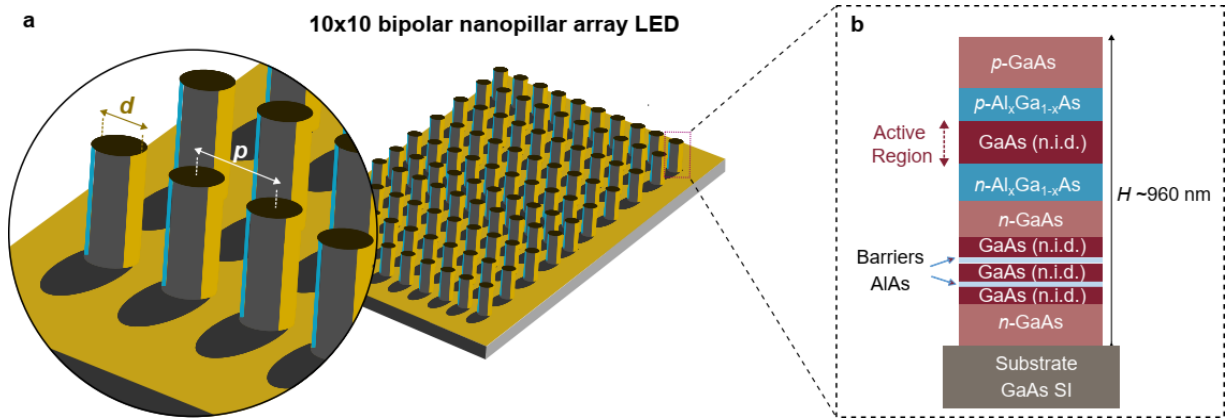

*Figure 1: III–V nanopillar array LED. (a) Schematic of a 10×10 nanopillar array LED. Inset (left) shows a magnified view detailing nanopillar diameter (d) and pitch (p). (b) Epilayer stack schematic of the GaAs/AlGaAs-based p-i-n nanopillar LED grown on a semi-insulating (SI) GaAs substrate.*

Devices where fabricated in a 10×10 square nanopillar array with pillar diameter (*d*) of 440 nm and pitch (*p*) of 1.3 μm (additional pillar sizes ranging from 440 nm to 870 nm, with pitches from 1.2 μm to 1.7 μm, respectively, were also fabricated). The fabrication used a top-down approach employing e-beam lithography and reactive ion etching (RIE), following our previously published process.[22] After nanopillar etching (~ 960 nm depth), samples were deoxidized using an ammonium hydroxide solution, then surface passivated using ammonium sulphide. Immediately afterward, a 100 nm thick $Si_xN_y$ dielectric layer was deposited by low frequency (380 kHz) PECVD, enhancing surface passivation via hydrogen ion bombardment,[3] and providing electrical isolation. We note in this process, no rinsing with water was used to preserve the sulphide layer formed in the GaAs surface. The sample was cleaned only using $N_2$. After the treatment, the sample was immediately transported to the PECVD deposition load lock chamber and pumped in vacuum conditions for PECVD deposition of low frequency $Si_xN_y$ layer. In all our tests, the typically time for air exposure before the dielectric coating was less than 5 min. Electrical contacts were realized by forming top via openings through the $Si_xN_y$ layer using a planarized etch back procedure, followed by metal sputter deposition. To enable light extraction and electroluminescence characterization, the top contacts were deposited at an angle following a previously described shadow-deposition procedure.[23] Figure 2a illustrates the angular metal deposition, where samples were positioned on one side of a triangular prism to achieve angled sputtering, creating metal-shadow region on nanopillar sidewalls (Figure 2b and 2c), clearly seen in the SEM image (Figure 2d). The shadowed regions facilitate light extraction from the nanopillar sides. The shadow size depends on the array pitch, metal deposition angle, and pillar height. The full fabrication details are provided in Supporting Information.

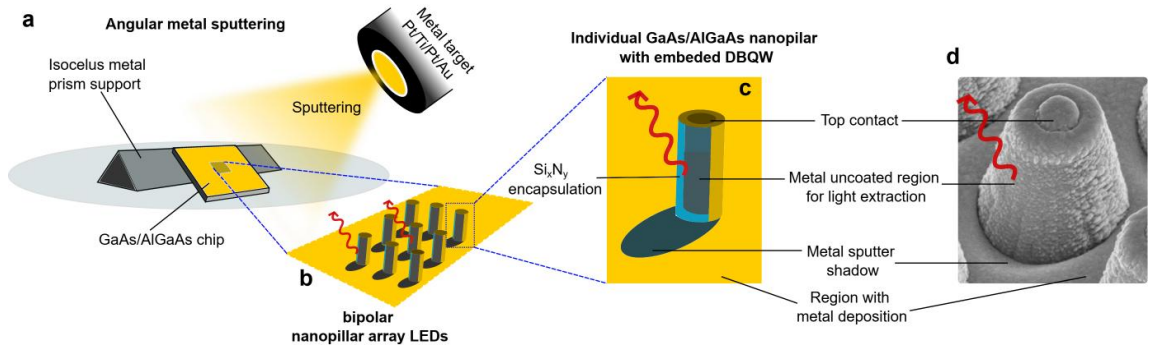

*Figure 2: Schematic of angular metal deposition to enable light extraction from nanopillar sidewalls. (a) The nanopillar LED chip is positioned on an isosceles metal prism at ~60° relative to the confocal metal sputtering target. (b) Schematic of the resulting metal-coated nanopillar array. (c) Detail of an individual nanopillar encapsulated by $Si_xN_y$ layer, showing the shadowed region with reduced metal coverage at the pillar base, enabling sidewall light extraction. (d) Scanning electron microscope (SEM) image of a metal-coated nanopillar after angular metal sputtering deposition.*

Figure 3 shows a fabricated 10×10 nanopillar array LED with a pitch size of 1.3 μm and nanopillar diameter of 440 nm. The devices feature electrical contact pads in a ground-signal-ground (G-S-G) configuration, enabling high-speed electrical characterization. For comparison, we also fabricated reference microLED devices consisting of single micropillars (~10 μm diameter) using standard $SiO_2$ passivation deposited by high-frequency PECVD, following a previously reported procedure.[24]

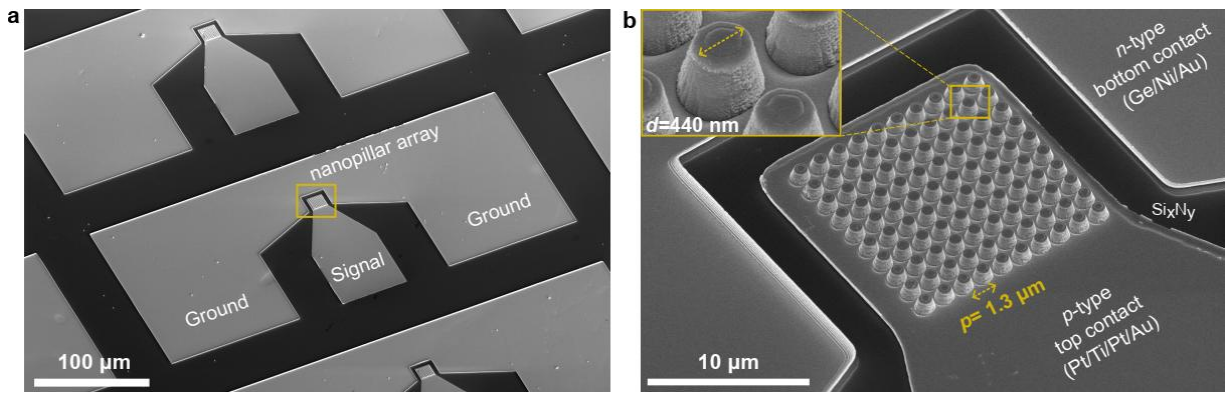

*Figure 3: Scanning electron microscope (SEM) images of fabricated GaAs-based nanopillar array LED devices. (a) SEM overview showing electrical contact pads arranged in a ground-signal-ground (G-S-G) configuration. (b) SEM image of a fabricated array nanoLED, with nanopillar diameter d=440 nm (inset) and pitch p=1.3 μm. The device includes n-type bottom contacts (Ge/Ni/Au), p-type top contacts (Pt/Ti/Pt/Au alloy), and a $Si_xN_y$ encapsulation layer acting as a passivating and insulating material.*

**Results and discussion**

**Electroluminescence and static characteristics**

The static characteristics of the nanopillar array LEDs were measured at room temperature using a ground-signal-ground (G-S-G) electrical probe. The DC bias was provided with a source meter (Keithley 2280S) via a bias-T (Mini-Circuits ZFBT-4R2GW+). The electroluminescence (EL) spectra were acquired using a multimode lensed fiber (Thorlabs LFM1F-1, NA=0.2, spot size ~25 μm) connected to a fiber spectrometer (Avantes AvaSpecHSC1024x58TEC-EVO-new). Figure 4a shows the measured EL spectra of a 10×10 nanopillar array LED (d=440 nm) as a function of the bias current using an applied voltage ranging from 2 V to 3.2 V. The broad emission peaking at $\lambda_{peak}$~861.1 nm corresponds to the typical emission from the band-edge transition of the intrinsic GaAs active region layer, at room temperature. Identical emission spectra were obtained for fabricated microLEDs consisting of single micropillars with a diameter ~10 μm (see Supporting Information). We observed the full width half maximum (FWHM) of the spectra increases monotonously with the operating bias current density for both nano- and microLEDs (see Supporting Information), with the microLEDs showing typically larger FWHM values (>32 nm) than the nanoLEDs (<26 nm). The nanoLED array example reported here was operated typically at lower current densities (≤1 kA/cm²) as compared to the microLED (>1 kA/cm²) to mitigate potential failures due to higher applied voltage. At these higher current densities, the microLED exhibits stronger band filling and increased self-heating, which broaden the emission spectrum, resulting in larger FWHM values (>32 nm), see Supporting Information S3, Figure S4. The fringe-like features in the spectra of Figure 4a are related to second-order effects from the spectrometer grating, which can be reduced by using an order-sorting filter. Additional measurements were carried out using a different commercial spectrometer (OceanInsight, model HR-4XR500-25), in which the fringes were not observed (spectra not shown). This confirms that the effect arises from the measurement setup and not from the nanoarray LEDs. Figure 4b shows the current-voltage (I-V) and the light-current (L-I) characteristics of the respective nanoLED with a turn-on voltage of 1.8 V, with the diode exhibiting a series resistance of 5.94 kΩ. The peak emission and turn-on voltage were consistent across arrays for a wide range of nanopillar sizes (see Supporting Information, S3). The measured output power of the nanoLED arrays shows operation at several tens of nW, reaching up to ~40 nW (Supporting Information, Figure S5). We note, however, these power levels, although higher than the pW-nW values typically reported in previous nanoLEDs, are significantly limited by the low numerical aperture (~0.2) of the lensed fiber used in our setup, the non-optimized collection angle of the fiber positioner fixed at ~15° from the normal axis, and additional optical losses from the metal-coated nanopillars, as discussed in Section Internal and external quantum efficiency.

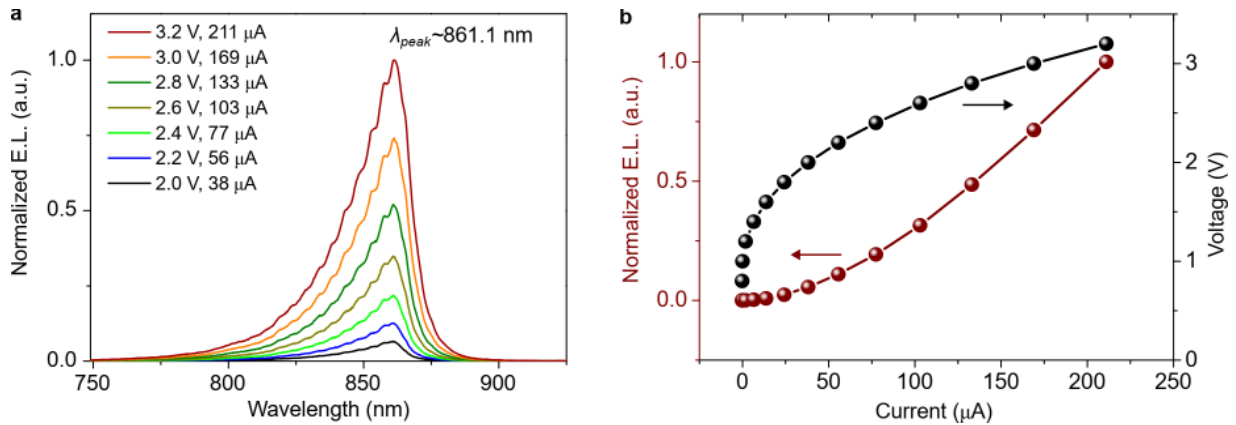

*Figure 4: Electroluminescence and static characteristics of nanopillar array LEDs. (a) Electroluminescence (EL) spectra of a 10×10 nanopillar array LED (d=440 nm) under various bias conditions. The main emission peak at $\lambda_{peak}$~861.1 nm corresponds to the intrinsic GaAs active region. (b) Light-current-voltage (L-I-V) static characteristics.*

**Dynamic response: time-resolved electroluminescence**

We have studied the high-speed modulation response of nanopillar array LED devices using TREL measurements to estimate the differential carrier lifetimes using a custom micro-EL (µEL) setup built upon a time-correlated single-photon counting (TCSPC) for high temporal precision as depicted in Figure 5a. The nanoLEDs were driven by electrical square-wave voltage pulses with peak-to-peak-voltage $V_{pp}$=1 V (-0.5 V to 0.5 V), pulse width $t_{in}$=10 ns, and repetition frequency $f_{in}$=1 MHz) using a pulse generator (Active Technologies, PG1072). The pulses were injected via the RF port of a high-bandwidth bias-T, with DC bias provided by a source meter (Keithley, 2280S). The modulated emission from the nanopillar array LED was collected by a lensed fiber (as described previously), and fed into a single-photon counting avalanche photodetector (APD, MPD PSM series). The APD was connected to a TCPSPC card (SPC-150N, Becker & Hickl),[24] which correlates photon arrival times at the APD (start signal) with the signal arrival times of the pulse generator (stop signal).[24] Photon arrival times are then binned to obtain a histogram which provides the time-dependent output intensity profile of the electrically modulated nanoLEDs. Figure 5b shows the TREL traces obtained for a 10×10 nanopillar array LED ($d$=440 nm) as a function of bias voltage. Additional measurements for nanoLEDs with larger pillar diameters are shown in the Supporting Information (Figure S6). The differential carrier lifetimes were estimated from the TREL measurements using a mono-exponential decay fitting function in *OriginLab* software, defined as follows:

$$N_{ph} = N_a e^{-\left(\frac{t-t0}{\tau}\right)} + N_0 \quad (1)$$

Where $N_{ph}$ is the normalized photon count, $N_a$ is the amplitude, $N_0$ is the offset value, $\tau$ is the differential carrier lifetime, and $t$ and $t_0$ are the measured and pulse-off times, respectively. Under low injection conditions, Figure 5b(i), we obtained a differential carrier lifetime of $\tau$=414 ± 11 ps. We note this value of differential carrier lifetime is only 2.2-fold shorter than the typically lifetimes obtained previously using optically-pumped nanopillars of similar sizes and with identical etching and surface passivation methods.[3] In our previous work, we have systematically compared $SiO_x$ and $SiN_x$ passivation, showing also that unpassivated nanopillars could not be experimentally resolved due to the extremely short lifetimes of the smaller size unpassivated pillars (e.g., ~150 ps for a 3 μm pillar[3]), and given the limited time resolution of our fastest detectors (~50 ps). Therefore, here we have compared our results with the case of microLEDs with ~10 μm diameter passivated pillar using sulfurization followed by $SiO_2$ coating, a passivation method that has been shown to provide a less effective passivation effect,[3] but providing a sufficiently long lifetime >1 ns at these large sizes. The differential carrier lifetime for the microLED at sub-mA current injection (1.5 V and 0.96 mA) is ~1.67 ns (see Supporting Information, Figure S7), which is only 4-fold longer than for the case of the nanoLED presented in Figure 5b. We note that the carrier lifetimes measured here are neither limited by the *RC* time constant of the diode circuit, $\tau_{RC}$, nor by the response time of each component of our TREL system (see table S.2 in Supporting Information). Lastly, as shown in Fig 5b(ii),(iii), by increasing the injection conditions the lifetime increases to values >600 ps. Similar increase of the lifetime for moderate injection conditions is observed for nanoLEDs with other pillar sizes (Figure S6 in Supporting Information), indicating reproducibility of the passivation method. The increase in lifetime with injection is attributed to the saturation of surface non-radiative states and to the fact that radiative recombination ($R_{rad} \propto Bn^2$, where *B* is the bimolecular recombination) grows faster with carrier density, *n*, than non-radiative recombination ($\propto n$), making radiative processes dominant at higher injection levels. Lastly, since the change in lifetime due to surface non-radiative effects are also size-dependent, Figure S8 provides additional TREL measurements confirming that nanopillar LEDs of increasing diameters exhibit longer lifetimes under equivalent current densities. At a current density of ~$10^4$ A/cm², the devices exhibit a lifetime increase from ~414 ps ($d$=440 nm) to 611 ps ($d$=750 nm), and at a current density ~$10^5$ A/cm², we observe a lifetime increase from ~465 ps ($d$=440 nm) to 706 ps ($d$=870 nm), confirming the trend of longer lifetimes for larger nanopillar sizes.

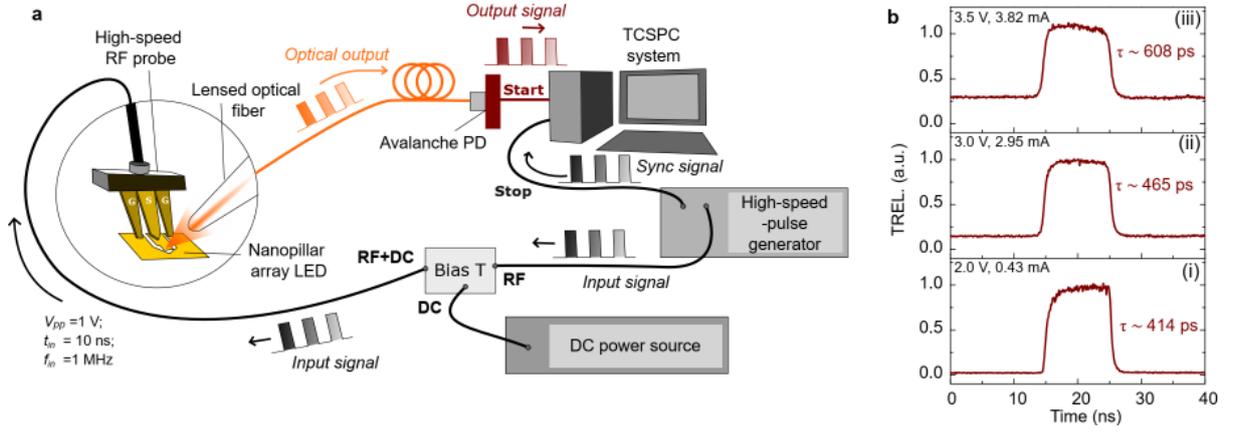

*Figure 5: Temporal electro-optical response of nanopillar array LEDs. (a) Schematic setup for the electro-optical modulation. The nanopillar LED array is electrically modulated using a high-speed ground-signal-ground (G-S-G) electrical probe. Emission from the nanoLED array is collected by a lensed fiber and detected by an avalanche photodiode (APD). Photon signals are recorded in time bins using a time-correlated single photon counting (TCSPC) system synchronized with the input electrical pulses from a high-speed pulse generator. (b) Time-resolved electroluminescence (EL) from a nanopillar LED array (d=440 nm) at forward bias voltage of (i) 2 V, (ii) 3 V and (iii) 3.5 V, driven by electrical pulses ($V_{pp}$=1 V, $t_{in}$=10 ns). The differential carrier lifetimes (τ) are also indicated.*

**Surface recombination velocity**

To estimate the surface recombination velocity (*S*) for our passivated nanopillar array LED devices, we applied the standard *ABC* rate equation model [19,21]. The differential carrier lifetime (τ) is related to the carrier density ($N_d$) through the steady-state rate equation:

$$\frac{1}{\tau} = A + BN_d + CN_d^2 \quad (2)$$

Where *B*=1.8×10⁻¹⁰ cm³ s⁻¹ is the bimolecular recombination coefficient [21], and *C*=3.5×10⁻³⁰ cm⁶ s⁻¹ is the Auger recombination coefficient [24]. The coefficient *A* is the surface recombination rate, and for a nanopillar with diameter *d*, and active material length $l_a$, *A* is proportional to the ratio of the surface area, σ=π$dl_a$, and the volume $V_a = \frac{\pi}{4} d^2 l_a$ of the active material, and is given by:

$$A = \frac{\sigma S}{V_a} = \frac{4S}{d} \quad (3)$$

Where *S* is the surface recombination velocity. In Figure 6 we plot *S* for the nanopillar LED device (*d*=440 nm), where we considered the estimated values of differential carrier lifetimes (τ) for carrier densities ranging from $N_d$=1×10¹⁶ cm⁻³ to 5×10¹⁸ cm⁻³ (similar to the doping ranges of the nanoLED epilayer, see Supporting Information S1). In Figure 6, we highlight in orange the lifetimes, τ, measured for our devices (shown in Figure 5b). For this range, we estimate a value of *S* for our nanoLED devices spanning from 0.7 × 10⁴ cm/s–2.7 × 10⁴ cm/s (highlighted in yellow). These results are in line with previous time-resolved PL studies [3], which revealed *S*~1.1 × 10⁴ cm/s for optically-pumped nanopillar GaAs/AlGaAs devices passivated by thin films $Si_xN_y$ deposited by LF-PECVD.

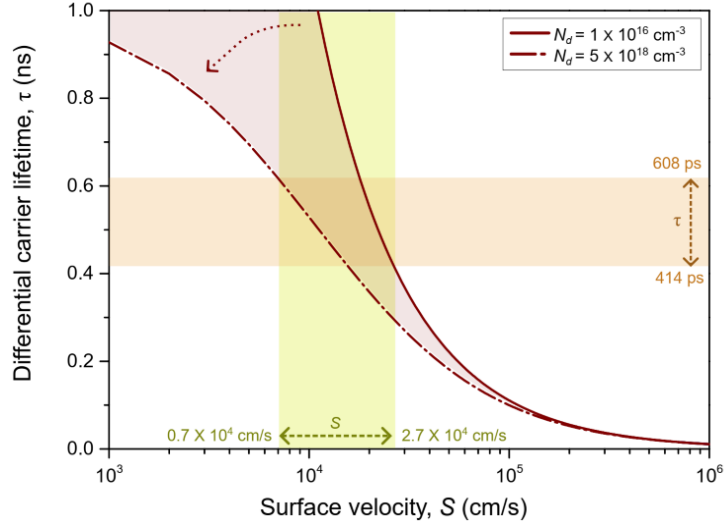

*Figure 6: Differential carrier lifetime (τ) as a function of surface recombination velocity (S) for various carrier densities. The solid and dashed-point brown lines show the differential carrier lifetimes at carrier density of $N_d=1\times10^{16}$ cm$^{-3}$ and $N_d=5\times10^{18}$ cm$^{-3}$, respectively. The orange highlighted region shows the range of the measured differential carrier lifetime by TREL. The yellow highlighted region shows the range of values for S for differential carrier lifetimes estimated and measured for the d=440 nm nanoLED device.*

**Modelling high-speed dynamic modulation of nanoLEDs**

We analyse the dynamic modulation properties of the nanoLED devices using a two-level semiconductor single-mode rate equations model, previously applied to III-V nanoLEDs [10,21] to describe the carrier population density ($N_d$) and photon density ($N_{ph}$). The model employed here was adapted from previous studies of metal–dielectric nanopillar cavities, including InP and GaAs-based metal-dielectric nanolasers,[2,25] and nanoLEDs.[10,21]

The carrier density ($N_d$) is given by:

$$\frac{dN_d}{dt} = \frac{\eta_I I(t)}{qV_a} - R_r - R_{nr}, \qquad (4)$$

and the photon number density ($N_{ph}$) is given by:

$$\frac{dN_{ph}}{dt} = R_{r,cav} - R_p. \qquad (5)$$

In equation 4, $R_r=\gamma B N_d^2$ represents the total spontaneous recombination rate (assuming the emission enhancement factor $\gamma=1$ for simplicity), and $R_{nr}=AN_d + CN_d^3$ represents the non-radiative recombination rate which accounts for surface and Auger recombination. Here, we assume the value of $S\sim1.1\times10^4$ cm/s which falls into the range estimated in the Figure 6. In equation 5, $R_{r,cav} = \gamma_m B N_d^2$ describes the radiative decay rate into the cavity mode, where $\gamma_m\sim0.01$ is the emission coupled to the cavity mode. In the present work we assumed a much smaller emission coupled to the cavity mode (~0.01) than in other reports of metal-dielectric coated nanopillars,[10,21] consistent with previous simulation results for non-ideal, non-optimized nanopillar cavities, corresponding to the case where the optical mode volume is much larger than the active emission volume, as reported in.[2] The photon escape rate is given by $R_p=N_{ph}/\tau_p$, where $\tau_p=\frac{\lambda_{peak} Q}{2\pi c}$, is the photon lifetime. This is determined from the cavity quality factor $Q=\frac{\lambda_{peak}}{FWHM}\sim34$ (with wavelength $\lambda_{peak}\sim861.1$ nm, and full width at half maximum (FWHM) ~25 nm from the emission spectra, see Figure 4(a)). The active volume $V_a=\frac{\pi}{4}d^2 l_a$ is defined by the nanopillar diameter $d$=440 nm and intrinsic active region length $l_a$=280 nm. The modulated injection current $I(t)$ was modelled as:

$$I(t) = I_b + \frac{V_{amp}}{nR} e^{-\left(4\log(2)\left(\frac{t-t_0}{\delta}\right)^2\right)} \quad (6)$$

Where $I_b$ is the injection current per nanopillar in the array, $V_{amp}$=0.5 V is the Gaussian pulse voltage amplitude, $R$=50 Ω is the impedance of the system (Supporting Information S5), $n$=100 pillars (10×10 array), $t_0$ is the center of impulse modulation, and $\delta$=0.1 ns is the pulse duration. Using a carrier injection efficiency ($\eta_I$) as a fitting parameter, the value $\eta_I$=0.75 provided the best fit with experimental data.[10] Figure 7 compares experimental TREL measurements for a 10×10 nanopillar array ($d$=440 nm) with simulations under various electrical pumping conditions. The dynamic pumping conditions for each nanopillar LED are estimated by assuming a uniform distribution of current per nanopillar ($I_{pillar}$=$I_{array}/n$). As the bias increased from 2.0 V ($I_{pillar}$=4.3 µA) to 3.5 V ($I_{pillar}$=38.2 µA), carrier lifetime increased from 414 ps to 608 ps, consistent with a transition toward the spontaneous emission bimolecular recombination regime. The dynamic model (equations 4 and 5) agrees with the experimental results, predicting lifetimes from 449 ps to 625 ps, and assuming injection current ranging from $I_b$~0.01 µA to 6.5 µA, respectively (identical to the current ranges per pillar experimentally obtained from the static I-V of the 440 nm nanopillar array LED). Based on the measured carrier lifetimes, we estimate a 3 dB-modulation bandwidth, $f_{3dB} = \frac{1}{2\pi\tau}$, ranging from $f_{3dB}$~0.26 GHz at moderate bias conditions (3.5 V, $I_{pillar}$=38.2 µA) up to $f_{3dB}$ ~0.38 GHz at low pumping conditions (2.0 V, $I_{pillar}$=4.3 µA). We note that due to the low extracted power (nW range) at the electrical pumping conditions evaluated here, the 3 dB bandwidth was estimated from lifetimes via time-correlated single-photon counting rather than direct small-signal modulation.

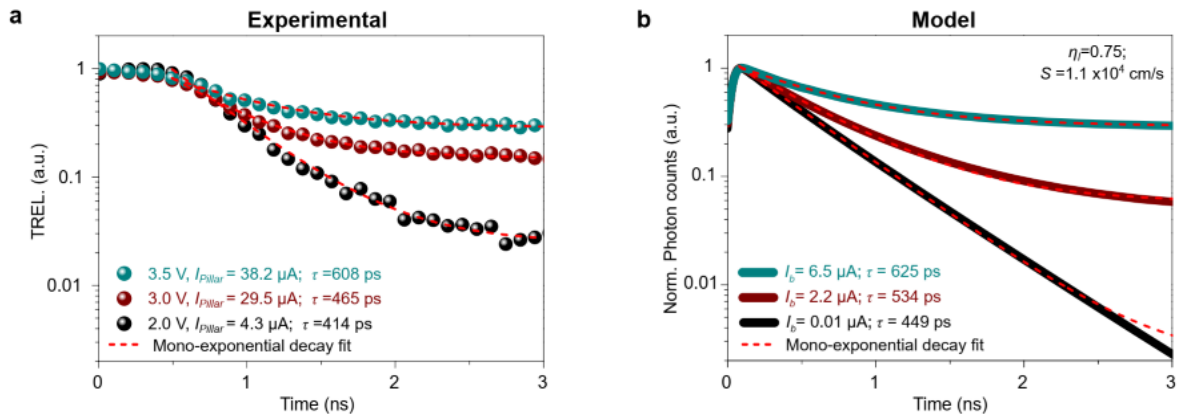

Figure 7: Experimental and simulated dynamic modulation of nanopillar LED arrays. (a) Experimental TREL measurements for a 10×10 nanopillar LED array (d = 440 nm) for various pumping conditions (electric current values are given per pillar). (b) Simulated dynamics based on the rate equation model for a single nanopillar LED (d=440 nm) under various injection bias current ($I_b$). Lifetimes were extracted using mono-exponential fitting. The experimental current levels reported in panel (a) include the AC modulation contribution, whereas in the model of panel (b) the values correspond only to the DC bias current.

**Internal and external quantum efficiency**

Using the previous analysis, we have estimated the IQE of our nanoLED devices. IQE is given by:

$$IQE = \eta_I \frac{R_r}{R_r+R_{nr}} = \eta_I \frac{BN}{A+BN+CN^2} \quad (7)$$

Where $\eta_I$=0.75 is the carrier injection efficiency taken from the model fitting of the dynamic modulation of nanoLEDs. We assumed surface recombination velocities of $S$~$1.1\times10^4$ cm/s for nanoLEDs, and $S$~$1.99\times10^5$ cm/s for microLEDs (value taken from *B. Jacob et al. (2023)*[3] for the case of microLEDs). Figure 8 shows the calculated IQE values for micro- and nanoLED devices. The IQE was calculated for a carrier density ranging from $N_d$=$10^{15}$ cm$^{-3}$ to $N_d$=$10^{20}$ cm$^{-3}$. At $N_d$=$3\times10^{17}$ cm$^{-3}$ (low pumping condition), the IQE is ~0.012 for nanoLEDs. In Figure 8, a range is highlighted from $N_d$=$10^{17}$ cm$^{-3}$ (moderate pumping conditions) to $N_d$=$3\times10^{19}$ cm$^{-3}$ (high pumping conditions). At $N_d$=$3\times10^{18}$ cm$^{-3}$ (moderate pumping condition), the IQE reaches up to ~0.26 for nanoLEDs compared to ~0.3 for microLEDs of similar architecture. Under high injection conditions the nanoLEDs can potentially exhibit IQE~0.45, limited only by Auger recombination and self-heating effects at high current density.

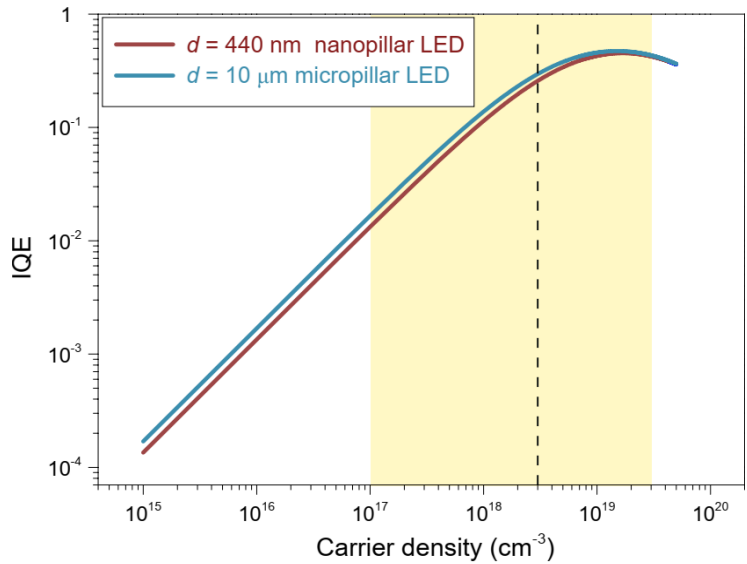

*Figure 8: Calculated internal quantum efficiency (IQE) as a function of carrier density for a nanoLED array (d=440 nm, solid brown line), and a microLED (d = 10 μm, solid cyan line). The yellow shaded region indicates the carrier density range analysed ($N_d$=$1\times10^{17}$–$3\times10^{19}$ cm$^{-3}$). A representative value of $N_d$=$3\times10^{18}$ cm$^{-3}$ (vertical dashed line), corresponding to moderate pumping conditions, is used for comparing IQE between micro- and nanoLED cases.*

Lastly, we have also estimated the EQE of our devices (see Supporting Information, Figures S9-S11 for full analysis). The EQE is given by:

$$\text{EQE} = \eta_c \text{IQE} \quad (8)$$

Where $\eta_c$ is the light extraction efficiency, and is determined by the numerical aperture (NA~0.2) of the lensed fiber used to collect the EL data (assuming an optimal collection angle from the sidewall pillars), ratio of extraction efficiency of the pillar ($\eta_{pillar}$) to the planar LED emission ($\eta_{bulk}$), and ratio of light coupled with metal coated and uncoated pillars ($\alpha$):

$$\eta_c = \frac{1}{4}\left(\frac{NA}{n}\right)^2 \left(\frac{\eta_{pillar}}{\eta_{bulk}}\right)\alpha \quad (9)$$

Where $n$≈3.55 is the refractive index of GaAs. An EQE ranging from $0.25 \times 10^{-4}$ to $8.4 \times 10^{-4}$ in the highlighted pumping conditions ($N_d$=$10^{17}$ cm$^{-3}$ to $N_d$=$3\times10^{19}$ cm$^{-3}$) was observed (See Supporting Information S6).

Table 1 summarizes the performance of the nanoLEDs of this work compared to previously reported III–V nanoscale light-emitting devices. In this work, the electrically pumped nanoLEDs are coated with metal layers for the electrical contacts. As a result, considering also the limited NA aperture of the lensed fiber and the non-optimized angle of the fiber positioner to couple light from the nanopillars, the estimated EQE for our devices (EQE<10$^{-3}$) is much lower than the reported IQE. We conclude that the main limitation for the EQE in our devices, besides potential limitations in injection carrier efficiency, is the low extraction efficiency due to the metal coating. This can be improved by either optimizing the angular metal deposition method (e.g., by increasing the pillar height, reducing the tapering effect, or by decreasing the pitch of the nanopillars as reported in [26]) or using indium tin oxide (ITO) transparent contacts. Despite this limitation, as shown in Table 1, the estimated EQE in this work 0.25–8.4 × 10$^{-4}$, is still a considerable improvement as compared to various reported electrically pumped architectures including *p-i-n* nanowire LED,[9] *p-i-n* nanopillar metal-dielectric LED,[10] *p-i-n* quantum dot PhC LED,[8] and *n-i-n* unipolar nanoLEDs.[22] The results are also comparable to the best results shown in optically pumped GaAs nanowire LEDs.[27]

**Table 1:** Comparison of this work with state-of-the-art III–V nanoscale LEDs.

| Device type | Material of the active region (III-V) | Volume of the active region (cm$^3$) | IQE | EQE at 300 K | Light extraction efficiency $\eta_c$ | Modulation Speed (ps) |
|---|---|---|---|---|---|---|
| *p-i-n* nanowire LED [9] | InP/InAs (bulk) | 1.6 × 10$^{-13}$ | - | 10$^{-6}$ | - | 370 |
| *p-i-n* nanopillar metal-dielectric LED [10] | InGaAs (bulk) | 4.36 × 10$^{-14}$ | - | 10$^{-4}$ | 0.035 | 100 |
| *p-i-n* QD/ PhC LED [8] | InAs/GaAs (QD) | 5.94 × 10$^{-16}$ | - | 10$^{-5}$ | - | 10 |
| *n-i-n* unipolar nanopillar array LED [22] | GaAs (doped bulk) | 1.6 × 10$^{-14}$ | 0.02 | <10$^{-5}$ | 4.73 × 10$^{-4}$ | 300 |
| Nanowire LED [27] (optically pumped) | GaAs (bulk) | 3.69 × 10$^{-13}$ | 0.3 | 5 × 10$^{-3}$ | - | 440 |
| **This work** | **GaAs (bulk)** | **5.42 × 10$^{-14}$** | **0.01-0.45** | **0.25–8.4 × 10$^{-4}$** | **0.002** | **414** |

**Conclusion**

This work demonstrates a combined ammonium sulphide surface treatment and low-frequency PECVD Si$_x$N$_y$ encapsulation for passivating electrically pumped GaAs/AlGaAs nanopillar *p-i-n* LEDs. The devices operate in the near-infrared, and time-resolved electroluminescence measurements under low injection conditions show differential carrier lifetimes ranging from 0.41 to 0.61 ns for a pillar diameter of 440 nm, a record large lifetime for electrically driven III–V nanoLEDs of this size. The devices can operate with a 3-dB modulation bandwidth ranging from 0.26–0.38 GHz, with measured lifetimes only four times shorter than those of unpassivated 10 μm microLEDs, confirming strong suppression of non-radiative recombination. A record-low surface recombination velocity of ~0.7 × 10$^4$ cm/s–2.7 × 10$^4$ cm/s was extracted, consistent with previous passivation tests, showing reproducibility of the passivation treatment, since our approach for passivating nanopillar devices has been verified over several stages of our current and previous work, first in optically pumped,[3] and now in electrically pumped *p–i–n* nanoLEDs, with consistent preliminary results also observed in *n–i–n* unipolar nanoLED,[22] all showing reproducible trends in lifetime enhancement. While the use of ammonium sulphide in our passivation can be scalable to larger GaAs wafers, the use of it requires appropriate industrial protocols to mitigate safety and environmental concerns. However, in this work, the ammonium sulphide treatment was mainly required because the etched samples were removed from the ICP-RIE tool for profilometry prior to dielectric deposition. In future integrated process flow, the need for this wet treatment can be avoided by end-point detection to monitor the etch depth in situ, followed by direct dielectric deposition without air exposure, thereby eliminating surface oxidation and removing the need for a wet chemical treatment.

Lastly, the internal quantum efficiency (IQE) reported in this work can potentially reach ~0.26 at moderate pumping, which compares with IQE ~0.3 for microLEDs with identical epilayer structure. A peak IQE value of ~0.45 can be potentially reached, limited by Auger recombination effect and self-heating effects at high current density, while external quantum efficiency (EQE) values ranging from $0.25 \times 10^{-4}$–$8.4 \times 10^{-4}$ were estimated. While improvements in EQE are still required in future work, these results demonstrate that sub-nanosecond modulation speeds can be achieved without substantially compromising radiative efficiency, making these nanoLEDs promising candidates for compact photonic circuits, high-speed optical interconnects, AR/VR displays, and neuromorphic edge computing applications.

**Supporting Information**

Epilayer design; Fabrication; Static characteristics: Electroluminescence and L-I-V characteristics; Dynamic response: high-speed modulation response; Coplanar waveguide (CPW) design of high-speed electrical contacts; External quantum efficiency (EQE)

**Author Contributions**

The manuscript was written through contributions of all authors. All authors have given approval to the final version of the manuscript.

**Acknowledgments**

We acknowledge access and support by the Micro and Nanofabrication Facility (especially Joana Santos and José Fernandes) and the Nanophotonics and Bioimaging Facility at INL. We acknowledge Jérôme Borme (INL) for the e-beam exposure of samples, and Iwan Davies and Stuart Edwards for the epitaxial growth of the III-V material. We acknowledge Qusay Raghib Ali Al-Taai, Jue Wang, and Edward Wasige of the University of Glasgow for the annealing of samples and the discussions of the design of the coplanar waveguide transmission line-based electrical contacts. We acknowledge José Figueiredo, Universidade de Lisboa, for discussions on the epitaxial design, and Xabier Quintana, Universidad Politécnica de Madrid (UPM), for providing the metal prism for angular deposition for nanoLED metal top contacts. BJ acknowledges the support in frame of the PhD program in Electrical Engineering, Electronics and Automation at Carlos III University of Madrid, Department of Electronic Technology, Group of Displays and Photonic Applications, Avda. de la Universidad, 30, 28911, Leganes, Madrid, Spain.

**Funding Sources**

We acknowledge the financial support by European Union, H2020-FET-OPEN framework programme, Project 828841 – ChipAI, and Horizon Europe, project 101046790 – InsectNeuroNano, and Fundação para a Ciência e a Tecnologia (FCT) projects 2022.03392.PTDC – META-LED and ERC-PT-A.